\documentclass[aps,pra,twocolumn,showpacs,showkeys,reprint,prb]{revtex4-1}
\usepackage{graphics}
\usepackage{graphicx}
\usepackage{amsmath}
\usepackage{color}
\definecolor{rojo}{cmyk}{0.0,1.0,1.0,0.05}
\usepackage{soul}
\begin{document}

\title{Three-body interaction effects on the ground state of one-dimensional anyons}

\author{J. Arcila-Forero}
\author{R. Franco}
\author{J. Silva-Valencia}
\email{jsilvav@unal.edu.co}
\affiliation{Departamento de F\'{\i}sica, Universidad Nacional de Colombia, A. A. 5997 Bogot\'a, Colombia.}

\date{\today}

\begin{abstract}

We investigated a one-dimensional system of anyons that interact with each other under a local three-body term. Using a fractional
Jordan-Wigner transformation, we arrived at a modified Bose-Hubbard model, which exhibits gapped and gapless phases. We built the phase 
diagram of the system fixing the hopping parameter or the statistics, showing the evolution of the critical points, which were estimated  with  
von Neumann block entropy. A superfluid to Mott insulator quantum phase transition with one particle per site can be driven by the statistics or the 
interaction. Specifically, we show that for larger angles there is a finite critical value of the interaction at which the Mott phase appears. Also, we 
found that the critical angles increase with the hopping. Diverse gapless phases were observed away from the pseudo-fermion limit.
\end{abstract}

\maketitle

\section{Introduction}

Commonly, particles in quantum theory are  classified into two types, bosons and fermions, depending on their particle exchange characteristics. 
For two bosons, the wavefunction remains invariant under particle exchange, whereas the exchange of two fermions leads to a multiplicative factor -1 
in the wavefunction. Physicists have proposed a third class of particles with nontrivial exchange statistics, anyons, particles carrying fractional 
statistics that interpolate between bosons and fermions~\cite{Leinaas-NCB77,Wilckzek-PRL82,Haldane-PRL91}. For two anyons under particle exchange, 
the wave function acquires a fractional phase $e^{i\theta}$, giving rise to fractional statistics with $0<\theta <\pi$.  It was quickly found that 
relevant quantum phenomena can be explained in terms of these new particles, the fractional quantum Hall effect~\cite{Tsui-PRL82,
Laughlin-PRL83,Halperin-PRL84,Camino-PRB05} being the best known, and nowadays it is believed that anyons will be important in condensed matter 
physics and topological quantum computation~\cite{Kitaev-AP03,Nayak-RMP08,Alicea-NP11}.\par 
The study of anyons was restricted for many years to two-dimensional systems. However, with Haldane's definition of fractional statistics, it 
was generalized to arbitrary dimensions~\cite{Haldane-PRL91}. Anyons in one dimension (1D) have been widely studied, their commutation 
relations~\cite{Kundu-PRL99}, dispersion relations and the dependence between generalized exclusion statistics and the anyonic 
angle~\cite{Osterloh-JPA00,Batchelor-PRL06}, correlation functions in the low-momentum regime~\cite{Calabrese-PRB07}, the dynamical properties in a weak harmonic 
trap~\cite{Hao-PRA12}, the ground state properties of hard-core anyons \cite{Patu-JSM15}, and mixtures of anyons under an external 
trap \cite{Zinner-PRA15} having been found. Also, it has been observed that the excitations of some one-dimensional models can be explained in terms of 
anyons ~\cite{Vitoriano-PRL09,Girardeau-PRL06}.\par
The creation and manipulation of physical setups that display fractional statistics is a huge challenge for experimental physicists, although several
proposals have been made~\cite{Paredes-PRL01,Duan-PRL03,Jiang-NP08,Aguado-PRL08,Longhi-OL12,Zhang-PRB15}. One of the most popular approaches for 
observing fractional physics is to confine ultracold bosonic atoms in an optical lattice, whose requirements have been considerably changed and 
simplified recently~\cite{Keilmann-NC11,Greschner-PRL15,Cardarelli-PRA16,Strater-PRL16}. The anyon-Hubbard model, which takes into account the kinetic 
energy of the anyons and the local two-body repulsion between them, arises from the above proposals. The ground state and the phase diagrams as a 
function of the tunneling, the local interaction, and the statistics have been studied using mean field theory and the density matrix renormalization 
group method~\cite{Keilmann-NC11,Greschner-PRL15,Arcila-PRA16,Zhang-PRA17}. In others studies, the quasimomentum distribution~\cite{Tang-NJP15}, the 
one-body reduced density matrix in a harmonic potential~\cite{Marmorini-JSM2016}, and a nontrivial topological Haldane phase were 
explored~\cite{Lange-PRL17}.\par
In the experimental setups of cold atoms using optical lattices, it is possible to tune the tunneling parameter of particles in a lattice, the 
interaction strength through Feshbach resonance, the density of the particles, and the dimensionality, and for this reason they have emerged as unusual 
laboratories for the realization of boson- and Fermi-Hubbard models, as well as for observing phase transitions without the intrinsic uncertainty posed 
by materials~\cite{Duchon-Arxiv2012,Greiner-N02b}. Also, these setups have allowed the observation of many-body interaction 
effects~\cite{Will-N10,Ma-PRL11}, stimulating new experimental proposals and theoretical calculations about the physical properties of bosonic 
systems with relevant three-body interactions between particles~\cite{JSV-PRA11,JSV-EPJB12,Daley-PRA14,Sowinski-CEJP14}.\par 
An unexplored problem consists of considering delocalized anyons in a one-dimensional optical lattice under local three-body interactions. This problem 
 is interesting because the three-body interactions between spinless or spinor bosons cannot generate a Mott insulator state with one particle 
per site and also change the phase diagram. On the other hand, it has been shown that the anyonic statistics localize the particles, and this can 
induce a quantum phase transition in systems with two-body interactions. In the present paper, we study the interplay between the above phenomena 
and write a Hamiltonian with two terms: the local three-body interaction and the kinetic energy. Using a fractional Jordan-Wigner transformation, this 
Hamiltonian was mapped to a modified Bose-Hubbard Hamiltonian with the tunneling depending on the local density and local three-body interactions, 
which will be studied numerically. The ground state can be gapped or gapless, in accordance with our density matrix renormalization group results. 
For a global density of one particle per site, we show that the interplay of statistics and the three body interactions leads to a 
Mott insulator state away from the pseudo-fermion limit. Note that at the above limit a Mott insulator state was found in the absence of 
interactions~\cite{Greschner-PRL15}. We build two phase diagrams, the chemical potential as a function of the statistical angle for a fixed hopping 
parameter and the chemical potential versus the hopping for a fixed anyonic angle. Although a complete study of the superfluid regions is beyond the 
scope of this paper, our results indicate that for non-integer densities smaller than one, the ground state is a superfluid that does not depend on 
statistics. After that, a paired phase appears even at the pseudo-fermion limit, this being an important difference between two- and three-body results. 
Also, we obtained a superfluid phase for densities larger than one at the pseudo-fermion limit. We expect that these facts will be observed in the near 
future when recent proposals for creating an anyon system and obtaining a regime where the three-body interaction domain is realized. We would like 
to point out that a prior report on the interplay between many-body effects and anyonic statistics has been made~\cite{Dai-NP17}.\par 
Our study is organized in the following way: The anyon-Hubbard model with local three-body interactions is explained in Sec. \ref{ii}. Using the 
von Neumann block entropy and the excitation gap in sec. \ref{iii}, we distinguish the quantum phases of the model and draw the phase diagrams of 
the model. Finally, our main findings and final comments appear in Sec. \ref{iv}.

\subsection{\label{ii} Model}

In one dimension, the operators of creation ($a_j^{\dagger}$) and annihilation ($a_j$) of an anyon at site $j$ satisfy the following commutation 
relations:
\begin{eqnarray}
a_j a_k^{\dagger} - e^{-i\theta sgn(j-k)}a_k^{\dagger}a_j=\delta_{jk},\\\nonumber 
a_j a_k- e^{i\theta sgn(j-k)}a_ka_j=0, 
\label{crelan}
\end{eqnarray}
\noindent where $\theta$ denotes the statistical phase, and the sign function (multistep function) is $sgn(j-k)=\pm 1$ for $j>k$ and $j<k$, and 
$sgn(j-k)=0$ for $j=k$. Note that, two particles on the same site reproduce the ordinary bosonic commutations relations.\par 
The quantum phases of a set of soft-anyons is a subject of great interest at this moment, and they come from the interplay between the hopping of 
the carries throughout the lattice and the interactions. In the present paper, we consider repulsive local three-body interactions, and the 
Hamiltonian is given by
\begin{eqnarray}
\label{aHmH}
H&=&-t\sum_j^{L-1}\left(     a_j^{\dag} a_{j+1}  + h.c.   \right) \\\nonumber 
&&+ \hspace{0.2 cm}\frac{W}{6} \sum_j^{L} n_j(n_j - 1)(n_j - 2),
\end{eqnarray}
\noindent $t>0$ being the tunneling amplitude connecting two neighboring sites, $W$ the on-site interaction, $L$ the length of the 
lattice, and $n_j$ the number operator. The first term in the Hamiltonian (\ref{aHmH}) is the kinetic energy with strength $t$, and  
the second term stems from the short-range interaction between three particles. Our unit of energy will be $W=1$ unless otherwise specified.\par
The fractional version of the Jordan-Wigner transformation
\begin{equation}
a_j=b_j \hspace{0.12cm}exp\left( i\theta \sum_{i=1}^{j-1} n_i \right),
 \label{maping}
\end{equation}
\noindent defined by Keilmann {\it et al.} establishes an exact mapping between anyons and bosons in one-dimension~\cite{Keilmann-NC11}. Here, the 
operator $b_j$ describes spinless bosons, which satisfy $[b_j,b_i^{\dagger}]=\delta_{ji}$ and $[b_j,b_i]=0$. Note that the representation of the 
number operator in terms of anyon or boson operators is the same ($n_i=a_i^{\dagger}a_i=b_i^{\dagger}b_i$).\par
Applying the anyon-boson mapping (\ref{maping}) to the anyon-Hubbard Hamiltonian with three-body interaction (\ref{aHmH}), we obtain a Hamiltonian 
in terms of boson operators, which allows us to explore the ground state but also establishes a link between the anyon-Hubbard Hamiltonian and 
its possible experimental realization in cold atom setups. Therefore, the bosonic version of the anyon-Hubbard Hamiltonian with three-body interaction 
is given by:
\begin{eqnarray}
\label{mawtbi}
H&=&-t\sum_j^{L-1}\left(     b_j^{\dag} b_{j+1} e^{i\theta n_j} + h.c.   \right) \\\nonumber
&&+ \hspace{0.2 cm}  \frac{W}{6} \sum_j^{L} n_j(n_j - 1)(n_j - 2).\\\nonumber
\end{eqnarray}
In this way, the above Hamiltonian describes bosons with the tunneling depending on the local density $te^{i\theta n_j}$ for hopping processes from 
right to left ($j+1\rightarrow j$). All of the influence of the fractional statistics is concentrated in this term. If the target site
$j$ is unoccupied, the hopping amplitude is simply $t$. If it is occupied by one boson, the amplitude reads $te^{i\theta}$, for
two bosons $te^{i2 \theta}$, and so on.\\
\begin{figure}[t!]
\setlength{\fboxrule}{0.47 pt}
\includegraphics [width=0.51\textwidth]{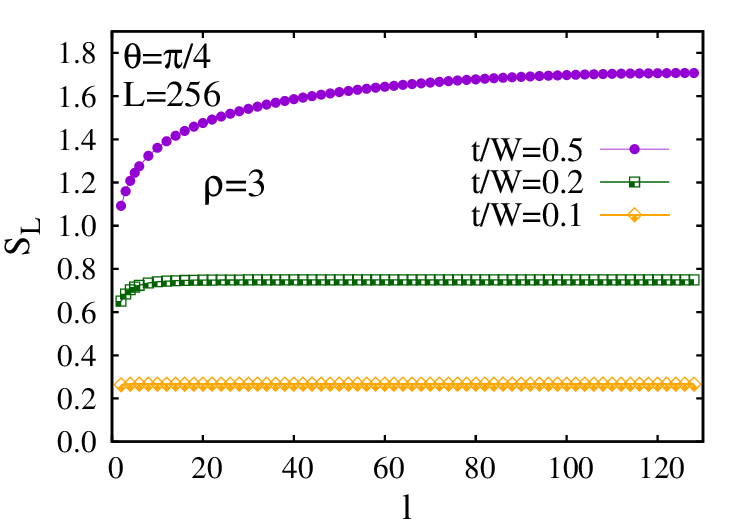} 
\caption{The von Neumann block entropy $S_L(l)$ as a function of $l$ for an anyon chain with $\theta=\pi/4$, $\rho=3$ and size $L=256$. 
 Here we consider three different values of the hopping parameter: $t/W=0.1, 0.2$, and $0.5$. Clearly, we can see that a change of state happens as 
 the hopping grows.}
\label{fig1}
\end{figure}
\subsection{\label{iii}  Results}

Nowadays it is well known that quantum information tools are useful for distinguishing different quantum phases, and also we can estimate the border 
or critical point that separates one region from another, although there is no  ``universal'' measure of entanglement that works for any quantum 
problem~\cite{Amico-RMP08}. Despite the relevance emphasized by many people and the possible applications of entanglement in diverse areas of physics, 
experiments for directly detecting and measuring the degree of entanglement of a system are challenging for science. Recently, in an experiment 
with atoms confined in optical lattices, it was possible to measure the quantum purity, R\'enyi entanglement entropy, and mutual information of a 
Bose-Hubbard system~\cite{Islam-N15}. These results reinforced the connection between quantum information tools and the description of degenerate 
gases realized in cold atom setups.\par 
Bipartite measures of entanglement are the most used. Therefore, we consider a system with $L$ sites divided into two parts. Part $A$ has $l$ sites 
($l=1,...,L$), and the rest form part $B$, with $L-l$ sites. The von Neumann block entropy of block $A$ is defined by 
$ S_A=-Tr\varrho_A  ln  \varrho_A$, where $\varrho_A=Tr_B \tilde{\varrho}$ is the reduced density matrix of block $A$ and 
$\tilde{\varrho}= |\Psi\rangle \langle \Psi|$ the pure-state density matrix of the whole system. For a system with open boundary conditions, the 
behavior of the von Neumann block entropy as a function of $l$ depends on the nature of the ground state and provides information about the type of 
phase, because it saturates (diverges) if the system is gapped (gapless)~\cite{Calabrese-JSM04}. Thus:
\begin{equation}
   S_L(l)= \left\{ \begin{array}{lcc}
             \frac{c}{6}ln [\frac{2 L}{\pi} sin (\frac{\pi l}{L})] + \Theta, & \text{critical}, \\
             \\ \frac{c}{6}ln[\xi_L]+\Theta^{\prime}, &\text{non critical},  
             \end{array}   
\right.
\label{lauchhh}
\end{equation}
\noindent $c$ being the central charge and $\xi_L$ the correlation length. The constants $\Theta$ and $\Theta^\prime$ are nonuniversal and model 
dependent.\par
\begin{figure}[t!]
\setlength{\fboxrule}{0.47 pt}
\includegraphics [width=0.49\textwidth]{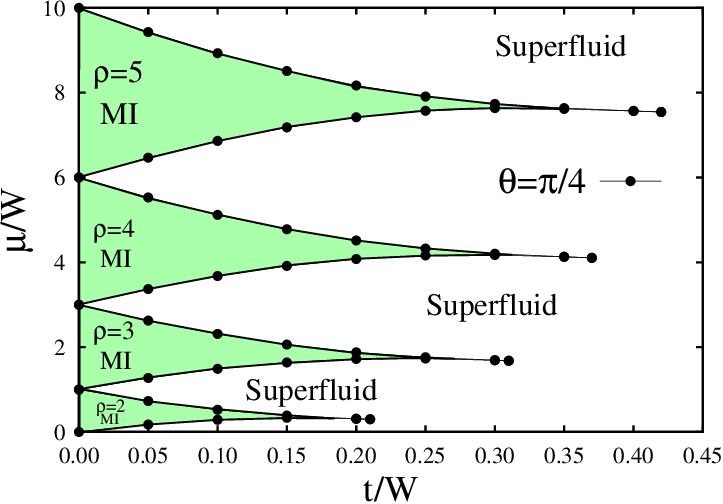} 
\caption{Phase diagram of the anyon-Hubbard model with local three-body interactions for $\theta=\pi/4$ and for the densities $\rho=2,3,4$ and $5$.
The points correspond to extrapolations to the thermodynamic limit from DMRG data, and the lines are visual guides. MI means Mott insulator regions.
The elongated form of the Mott insulator lobes suggests that the transitions are of the Kosterlitz-Thouless type, and the critical points can be calculated 
by means of gap-scaling analyses~\cite{Carrasquilla-PRA13,Dalmonte-PRB15} or a scaling of the Luttinger-liquid parameter~\cite{Clay-PRB99} or the 
L\"auchli and Kollath estimator~\cite{Lauchli-JSM08}.}
\label{fig2}
\end{figure}
To determine the ground-state wave function ($|\Psi\rangle$) of a lattice with $L$ sites and $N$ particles, we used the finite-size density matrix 
renormalization group algorithm (DMRG) with open boundary conditions, and in order to improve the calculations, we implemented the dynamical block 
state selection (DBSS) protocol based on a fixed truncation error of the subsystem’s reduced density matrix instead of using a fixed number of 
preserved states in the DMRG sweeps~\cite{Legeza-PRB03}. Also, we truncated the local Hilbert space by considering only $\rho+5$ states  when the 
density of the particles is $\rho=N/L$~\cite{Carrasquilla-PRA13}. With the above considerations, we obtained a discarded weight of around $10^{-10}$  
or less, and the maximum number of states maintained was $m = 1200$.\par
\begin{figure}[t!]
  \centering
\setlength{\fboxrule}{0.5 pt}
 \includegraphics [width=0.46 \textwidth]{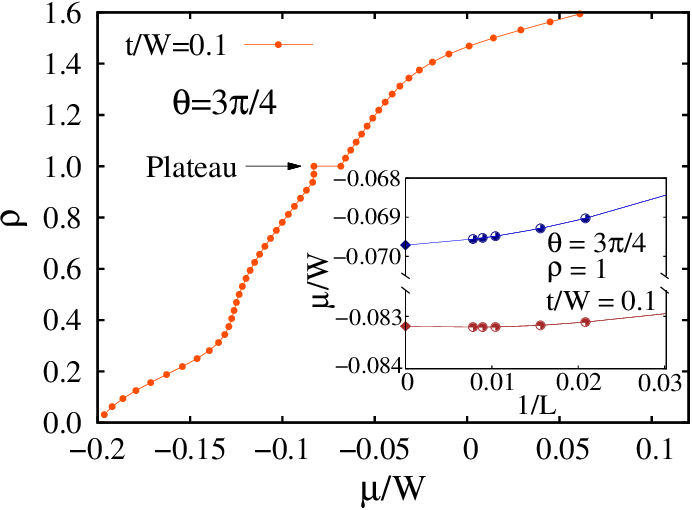} 
 \caption{ Global density $\rho$ versus the chemical potential $\mu/W$ for $t/W=0.1$ and a statistical angle $\theta=3\pi/4$. The system 
 exhibits a Mott plateau at integer density $\rho=1$. Inset: System size dependence on the chemical potential of anyons with three-body interactions 
 for statistical angle $\theta=3\pi/4$ and $\rho=1$. The upper set data  
 correspond to the particle excitation energy and the lower to the hole excitation energy (the lines are visual guides). 
 The values for $1/L=0$ (diamonds) correspond to an extrapolation to the thermodynamic limit.}
\label{fig3}
\end{figure}
The evolution of the von Neumann block entropy $S_L(l)$ as the size of the block increases is shown in Fig. \ref{fig1} for a lattice with global 
density $\rho=3$, statistical angle $\theta=\pi/4$, and three different values of the hopping: $t/W=0.1$, $t/W=0.2$, and $t/W=0.5$. For small values 
of the hopping parameter, we expected that the particles would tend to localize and the entanglement would be small. This happened for $t/W=0.1$, and 
we observed that the von Neumann block entropy saturates very quickly and has a small value. The above behavior continues for larger values of the 
hopping parameter ($t/W=0.2$), but the maximum numerical value is larger, indicating that the entanglement has increased. These results indicate that 
for a wide range of values of the hopping the system remains in a phase characterized by a finite correlation length, in accordance with the 
expression Eq. (\ref{lauchhh}). However, things change for larger values. For $t/W=0.5$, the von Neumann block entropy increases smoothly and tends 
to diverge with the block size, which characterizes a critical state. In this way, the calculation of the von Neumann block entropy allows us to 
distinguish between two quantum phases in the system (one critical and the other non-critical) and thus to discern the appearance of a phase 
transition as we increase the tunneling of the particles. This can be deduced from the fact that the entropy changes its behavior (saturate-diverge) when 
we increase the value of $t/W$. In this case, considering that the system density is an integer and that we are facing a non-critical phase, the 
results suggest that the system could be found to be in a Mott-insulator phase characterized by having a finite-correlation length. This would 
explain the fact that the block entropy increases and saturates rapidly at a constant value. On the contrary, in the critical region, the system 
could exhibit a superfluid state, characterized by a divergent behavior of the von  Neumann block entropy and the delocalization of the particles 
along the lattice. In conclusion, for a fixed global density $\rho$, the anyon-Hubbard model with local three-body interactions exhibits a Mott 
insulator to superfluid transition for a finite value of $t/W$.\par
\begin{figure}[t!]
  \centering
\setlength{\fboxrule}{0.54 pt}
 \includegraphics [width=0.48 \textwidth]{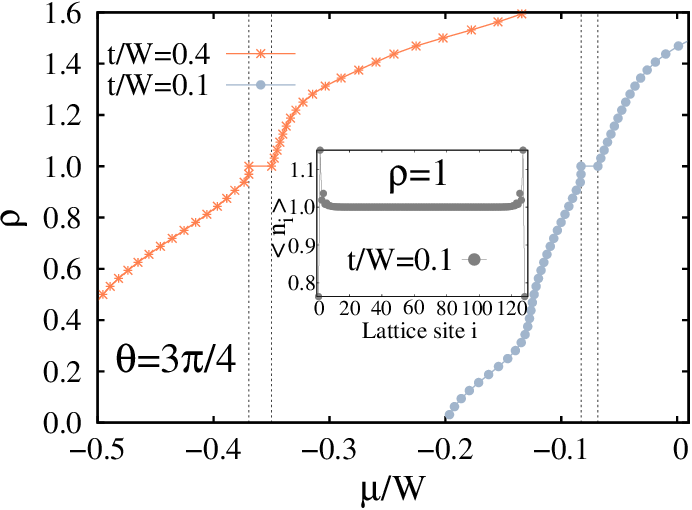} 
 \caption{ Global density $\rho$ as a function of the chemical potential $\mu/W$ at the thermodynamic limit. Here, we fix the statistical angle to 
 $\theta=3\pi/4$ and consider $t/W=0.1$ and $t/W=0.4$. The vertical dashed lines delimit the insulator phases. In the inset, we show the density 
 profile along the lattice for $\theta=3\pi/4$, $t/W=0.1$ and $L=120$.}
\label{fig4}
\end{figure}
The phase diagram of anyons in one dimension under two-body interactions shows Mott lobes surrounded by superfluid regions, with the Mott areas 
decreasing as the global density grows. In particular, it was found that the critical point for a statistical angle $\theta=\pi/4$ and a fixed global 
density $\rho=3$ is located at $t/U=0.172$, $U=1$ being the strength of the two-body interactions and the energy scale for this 
case~\cite{Keilmann-NC11, Arcila-PRA16}. In Fig. \ref{fig1}, we observe that the Mott lobe for $\rho=3$ survives for values greater than $t/W=0.2$, 
which indicates that the Mott lobes with three-body interactions will be larger than the lobes with two-body interactions, indicating that three-body 
interactions generate a larger localization in the system, in accordance with previous results for bosons with three-body interactions.\par
We found that the Hamiltonian problem (\ref{aHmH}) exhibits Mott and superfluid states, which are characterized by the presence or the lack of an 
excitation gap at the thermodynamic limit. For a lattice with finite size, the excitation gap is 
$\Delta\mu(L)=\mu^{p}(L)-\mu^{h}(L)=E_0 (L, N+1)+E_0 (L, N-1) - 2E_0 (L,N) $, where $E_0 (L, N)$ denotes the ground-state energy for $L$ sites and 
$N$ particles. In particular, a Mott-insulator phase is characterized by the presence of a positive gap at the thermodynamic limit, since the global 
density is an integer, while the superfluid phase is characterized by the fact that no gap exists at the thermodynamic limit. In Fig. \ref{fig2}, we 
show the thermodynamic limit values of the chemical potential of adding ($\mu^{p}$) and removing ($\mu^{h}$) a particle as a function of the hopping 
for a statistical angle $\theta=\pi/4$ and different global densities $\rho$. As is well known, the gapped regions are surrounded by superfluid 
(gapless) ones, and we obtained that upon increasing the global density, the insulator lobes increase and the position of the critical points moves 
to greater values of $t/W$ in a manner similar to the boson problem ($\theta=0$)~\cite{Avila-PLA14}, but for anyons the critical points are greater 
than for bosons. These results indicate that the three-body interaction and the statistics favor the localization of particles and that greater 
kinetic energy is required to delocalize the particles and generate superfluid states. Another characteristic of Fig. \ref{fig2} is the absence of 
the insulator region for $\rho=1$, which is due to the fact that the quantum fluctuations, having on average only one particle per site ($\rho=1$) 
and $\theta=\pi/4$, are insufficient for generating particle localization. The elongated shape of the Mott-insulator lobes indicates that the gap 
closes slowly and possibly follows a Kosterlitz-Thouless formula~\cite{Arcila-PRA16}. The length of the Mott regions on the vertical axis at $t/W=0$ 
varies with the density of the system. It is notable that for the Bose- and anyon-Hubbard models with two-body interaction, this length does not 
change with the density. Rather, it is constant and equal to one in the scale of $\mu/U$. When we consider three-body interactions, the length of the 
lobe with global density $\rho$ at $t/W=0$ is $\rho-1$ for both bosons and anyons with $\theta=\pi/4$.\par
\begin{figure}[t!]
\setlength{\fboxrule}{0.47 pt}
\includegraphics [width=0.51\textwidth]{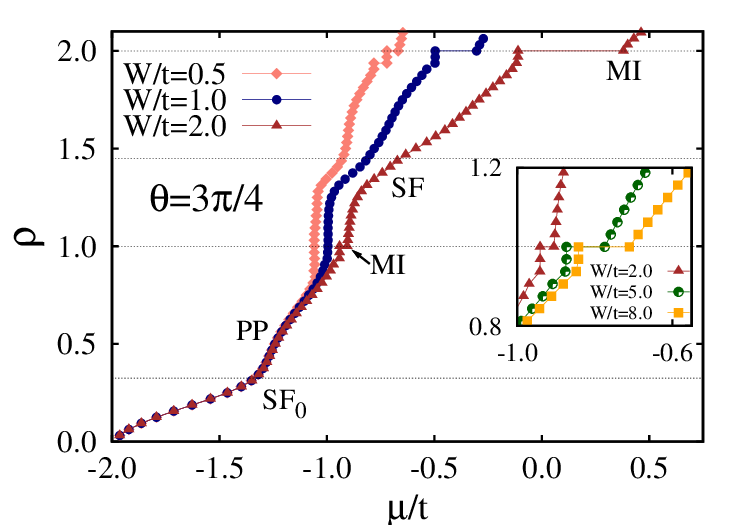} 
\caption{Global density $\rho$ versus the chemical potential $\mu/t$ at the thermodynamic limit with a statistical angle $\theta = 3\pi/4$ for 
different values $W=0.5$, $1.0$, and $2.0$. Here the unit of energy is the hopping parameter $t=1$. A zoom of the main figure around the first plateau 
appears in the inset, where the three-body local interaction is $W=2.0$, $5.0$, and $8.0$. The horizontal lines are visual guides.}
\label{fig5}
\end{figure}
One of the main findings in the anyon-Hubbard model is that the statistics favor the localization of the particles, which is reflected in the 
increase in the area of the Mott lobes as the statistical angle $\theta$ tends toward $\pi$~\cite{Keilmann-NC11}. The absence of the $\rho=1$ Mott 
insulator lobe in spinless or spinor bosons chains under local three-body interactions is a very well-known result, but in the present paper, our 
subject of study consists of anyons, and we expected that the statistical angle would generate new findings. Hence we wanted to explore if for 
$\theta>\pi/4$ new insulator phases can appear. Taking into account the above and the fact that a Mott insulator state appears in the absence of 
interactions at the pseudo-fermion limit ($\theta=\pi$)~\cite{Greschner-PRL15}, we decided to distance ourselves from the above limit. Therefore, 
we fixed the statistical angle to $\theta=3\pi/4$ and studied the evolution 
of the chemical potential as the number of anyons increases. This leads us to Fig. \ref{fig3}, where we show the global density as a function of the 
chemical potential at the thermodynamic limit. Note that as we increase the overall density of the system, the chemical potential increases for all 
the non-integer densities, but for a global density of $\rho=1$, a plateau appears, which indicates the existence of a finite gap at the thermodynamic 
limit for this  density, given by the width of the plateau. In the inset of Fig. \ref{fig3}, we show the evolution of the energies  for adding 
($\mu^{p}$) and removing ($\mu^{h}$) particles versus the inverse of the lattice size for anyons with $\theta=3\pi/4$ and density $\rho=1$. This 
evolution is quadratic for $t/W=0.1$, and at the thermodynamic limit we obtain $\Delta\mu/W=lim_{L,N\rightarrow \infty} [\mu^p(L)-\mu^h(L) ]=0.0135$, 
which corresponds to the width of the plateau and allows us to conclude that the ground state for $\theta=3\pi/4$ and $\rho=1$ is an insulator.\par
\begin{figure}[t!]
\setlength{\fboxrule}{0.47 pt}
\includegraphics [width=0.5\textwidth]{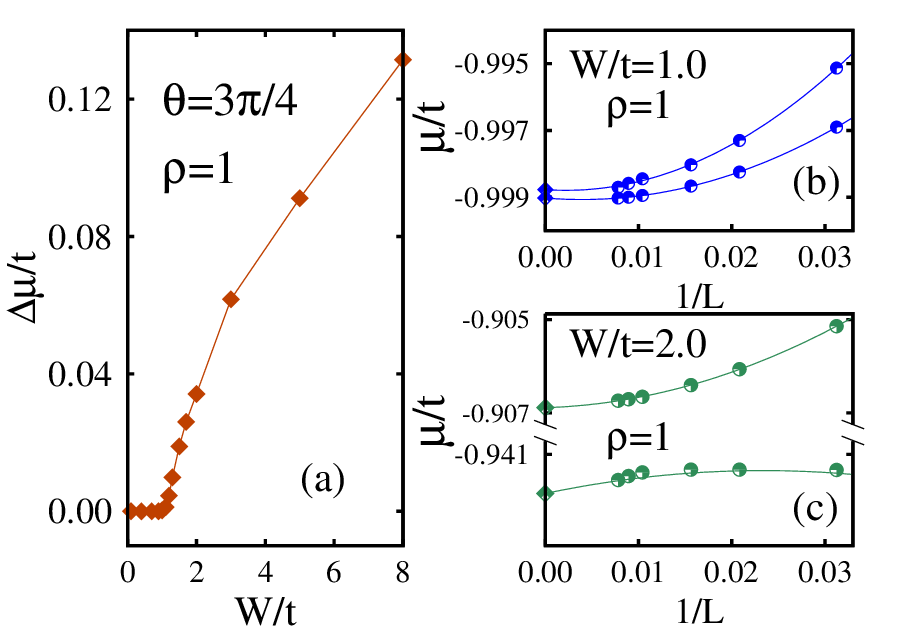} 
\caption{Left panel: The excitation gap at the thermodynamic limit as a function of the three-body interaction for a system of anyons with 
$\theta=3\pi/4$ and global density $\rho=1$. The critical point $W_c/t=1.1$ separates a gapless region and a Mott insulator one. Right panel: 
System size dependence on the chemical potential of a system with one particle per site, $\theta=3\pi/4$, and local interaction $W/t=1.0$ and 
$W/t=2.0$ for Figures (b) and  (c), respectively. In both figures, the upper set of data corresponds to the particle excitation
energy and the lower one to the hole excitation energy. The points are the DMRG results, and the lines are visual guides. The values for $1/L = 0$
correspond to an extrapolation to the thermodynamic limit, using a second-order polynomial in $1/L$.}
\label{fig6}
\end{figure}
Fixing the hopping parameter to $t/W=0.1$, we can compare Figs. \ref{fig2} and \ref{fig3} and observe that the quantum fluctuations are very small for 
$\theta=\pi/4$. Therefore, the interaction term is unimportant and the ground state is a superfluid (gapless), and the first insulator state is 
obtained for a global density of $\rho=2$. However, as the statistical angle increases, the particles are more localized, the effect of the effective 
interaction term grows, and an insulator state is obtained for $\rho=1$. The above insulator state is surrounded by gapless regions (see 
Fig. \ref{fig3}), but we note that the evolution is different for values greater than or less than $\rho=1$, and also for $\rho<1$ two diverse 
regions can be identified, which indicates that the increase of the statistical angle enriches the superfluid regions.\par 
The density profile of the particles along a lattice of $L=120$ with $\theta=3\pi/4$ and $t/W=0.1$ is shown in the inset of Fig. \ref{fig4}. 
At the ends of the lattice, we note strong fluctuations due to the open 
boundary conditions considered in our study. Despite this, we obtain that at each site there is one particle, i. e. $<n_i>=1$. This result, as well as 
the fact that there is a finite gap at the thermodynamic limit, allows us to conclude that the ground state is a Mott insulator, which is generated 
by the interplay of statistics and interactions, because for spinless and spinor bosons under local three-body interactions it is impossible to 
obtain a Mott insulator lobe with $\rho=1$. To distinguish the Mott regions from the superfluid ones, we draw vertical lines in Fig. \ref{fig4}, 
which shows the global density as a function of the chemical potential at the thermodynamic limit for $\theta=3\pi/4$ and two different values of the 
hopping $t/W=0.1$ and $t/W=0.4$. From this figure it is clear that a quantum phase transition from a Mott insulator to a superfluid phase will take 
place, and we observe that for both values of $t/W$ the compressibility ($\kappa=\partial\rho/\partial\mu$) is always greater than zero, which 
indicates the absence of a first-order transition. Also, we obtained that the overall behavior of the global density as a function of $\mu/W$ is the 
same for both values of the hopping. However, the Mott gap is obviously larger at $t/W=0.4$ than at $t/W=0.1$. This unusual result can be understood 
by taking into account that for generating a Mott insulator state with $\rho=1$ under local three-body interactions, we need the interaction term 
to be important. For this, the quantum fluctuations must grow, which is caused by the hopping, and with the localization due to the statistics we 
obtain a larger effective interaction term, and finally the Mott gap increases with the hopping.\par
\begin{figure}[t!]
\setlength{\fboxrule}{0.47 pt}
\includegraphics [width=0.51\textwidth]{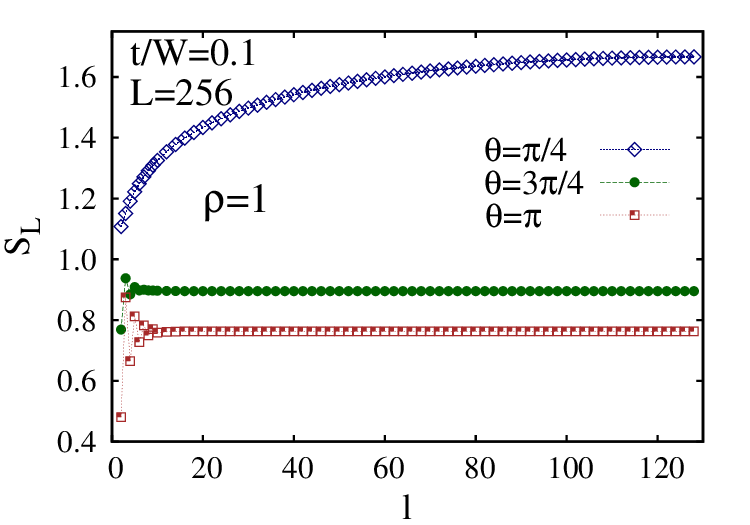} 
\caption{ The von Neumann block entropy $S_L(l)$ as a function of $l$ for a system with size $L=256$, $\rho=1$ and $t/W=0.1$. 
 Here we consider three different values of statistical angle: $\theta=\pi/4$, $\theta=3\pi/4$, and $\theta=\pi$. The statistical angle drives 
 a quantum phase transition.}
\label{fig7}
\end{figure}
Up to this point, we know that for zero and small values of the statistics ($\theta \leq\pi/4$) there is not a Mott insulator state with a global 
density $\rho=1$ whatever the value of the three-body interaction and that at the pseudo-fermion limit ($\theta=\pi$) there is a Mott insulator state 
in the absence of interactions. Also, our results away of the extreme limits suggest that both the statistics and the 
interaction are necessary to generate a Mott insulator state with one particle per site. To observe more clearly the effect of the three-body 
interaction, we temporarily change our energy unit to $t=1$ and express all quantities in terms of this. The global density as a function of the 
chemical potential $\mu/t$ appears in Fig. \ref{fig5}, where a statistical angle of $\theta=3\pi/4$ and different values of the local three-body 
interaction were considered. For global densities smaller than $\rho=1$, the before reported $SF_0$ superfluid  ($\rho < 0.34$) and the 
paired ($\rho > 0.34$) phases were obtained~\cite{Greschner-PRL15}, but the new fact observed is that the first phase and in some region of the 
second they do not depend of the three-body interaction, because there are few particles and the effective interaction is determinate by the 
statistics. We note that the effect of interactions is more striking from the global density $\rho=0.75$.
After the above density the equation of state $\rho(\mu)$ depends strongly on the interaction, note that the existence or not of plateaus depends on 
the value of the local interaction. For small values of interaction $W/t=0.5$ and $W/t=1.0$ there is not a plateau at a global density $\rho=1$ 
indicating that there is not a Mott insulator state for these values (see Fig. \ref{fig6}(b)), also we can see that the scope of the curve increases 
suggesting a separation of phases (PS), after this a paired phase and Mott insulator state with $\rho=2$ were obtained. When the three-body interaction 
increases ($W/t=2.0$) a plateau appears at the density $\rho=1$, where the evolution of the width of the plateau  as a function of the latiice size 
is shown in Fig. \ref{fig6}(c), being clear that at the thermodynamic limit the width of the plateau is finite and the ground state is a Mott 
insulator. Keeping the statistical angle and the value of the interaction, we obtain that between the Mott insulator states for the integer densities 
there is a superfluid phase ($SF_{\pi}$). From Fig. \ref{fig5}, we note that the ground state changes drastically as the three-body interaction 
increases, for non-integer densities the ground state goes from a paired phase to a superfluid one, at density $\rho=1$ a change from a gapless phase 
to a Mott insulator state takes place. Also, we can see that the width of the plateaus at the integer densities increases with the three-body 
interaction (see the inset of Fig. \ref{fig5}, where a zoom of the plateau at $\rho=1$ is shown for $W/t=2.0$, $5.0$, and $8.0$ ). In conclusion for 
a system of anyons with a statistical angle $\theta=3\pi/4$ without interactions there is not a Mott insulator state with one particle per site, once 
the interaction is turn on, the Mott insulator state appears from a critical value of the interaction. Figs. \ref{fig6}(b) and \ref{fig6}(c) tell us 
that the critical value of the interaction is between $W/t=1.0$ and $2.0$, where the width of the plateau vanishes and has a finite value 
respectively. The evolution of the width of the plateau as a function of the local interaction at the thermodynamic limit is shown in 
Fig. \ref{fig6}, where two different regions are clear, the first one is a gapless region for small values of  $W/t$ and a Mott insulator phase, 
where the gap increases monotonously with the three-body interaction. We estimate that the quantum critical point is  $W_c/t=1.1$. It is expected 
that the critical three-body interaction will depend on the value of the statistical angle $\theta$.\par 
\begin{figure}[t!]
\setlength{\fboxrule}{0.47 pt}
\includegraphics [width=0.505\textwidth]{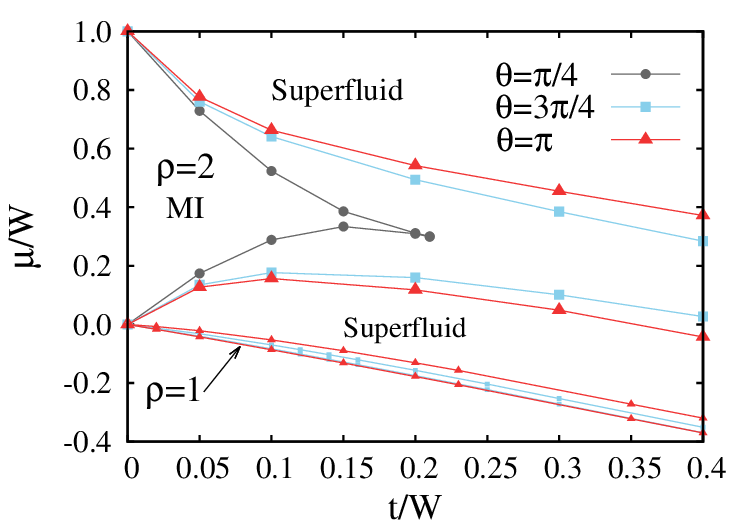} 
\caption{ Phase diagram of the anyon-Hubbard model with local three-body interactions for $\theta=\pi/4$, $3\pi/4$, and $\pi$. Mott lobes and a
superfluid phase were found, and their boundaries are marked by points that are extrapolations to the thermodynamic limit from DMRG data. The lines are 
visual guides.}
\label{fig8}
\end{figure}
Previously, we showed that the behavior of the block von Neumann entropy allows us to distinguish between critical and noncritical ground states, for 
instance superfluid and Mott insulator states. With respect to the appearance of insulator states for larger values of the statistical angle, we want 
to verify the above results by means of the calculation of block entropy (\ref{lauchhh}). We present the results in Fig. \ref{fig7} for three different 
values of the angle, $\theta=\pi/4$, $\theta=3\pi/4$, and $\theta=\pi$, setting the global density at $\rho=1$, the lattice size at $L=256$, and the 
tunneling at $t/W=0.1$, where again and untill the end of the paper our unit of energy will be $W=1$. For $\theta=\pi/4$, we observe that the entropy 
always increases until it diverges, which indicates that the system is in a superfluid phase. This is consistent with our previous results, in which 
a non-insulator region was found for this angle and this density (Fig. \ref{fig2}). On the other hand, the above behavior of the block entropy changes 
for larger values of $\theta$. Specifically, we observe that the block entropy increases very quickly exhibiting an oscillator behavior for small 
values of the block size, but the range increases with the angle. We cannot explain this oscillator behavior, which is due to the statistics. After 
the short oscillation, the block entropy remains constant, showing that the system has a finite correlation length. Hence the ground state is a Mott 
insulator one. Another interesting fact is that the overall value of the von Neumann block entropy diminishes as the statistical angle grows, which 
indicates that the entanglement decreases. The above discussion allows us to conclude that the statistical angle $\theta$ drives a quantum phase 
transition from a superfluid to a Mott-insulator phase with $\rho=1$.\par 
\begin{figure}[t!]
\setlength{\fboxrule}{0.45 pt}
\includegraphics [width=0.49\textwidth]{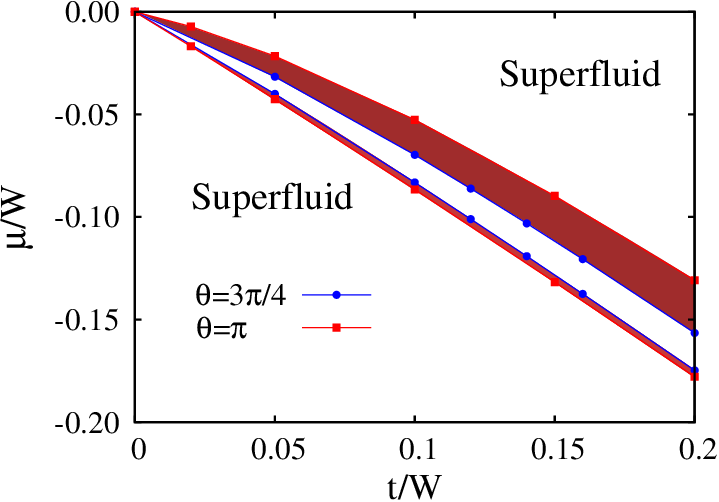} 
\caption{ The $\rho=1$ Mott lobe at the thermodynamic limit for a statistical angle $\theta=3\pi/4$ and $\theta=\pi$. The lines are visual guides.}
\label{fig9}
\end{figure}
We found that the statistics induce a Mott insulator region for larger values of the angle $\theta$. This fact suggests that the phase diagram changes 
with the statistical angle, and we have to calculate this for statistical angles greater than $\theta=\pi/4$. In Fig. \ref{fig8}, we also draw the 
results for $\theta=\pi/4$ shown in Fig. \ref{fig2}, in order to compare them with the new results obtained, $\theta=3\pi/4$ and $\theta=\pi$. We want 
to emphasize the absence of the $\rho=1$ Mott lobe for $\theta=\pi/4$. At the atomic limit ($t/W=0$), we observe that the Mott lobe regions are given 
by $\mu/W=\rho-1$ regardless of the statistical angle; hence there is no Mott lobe with $\rho=1$ at this limit. For the non-zero hopping parameter, we 
see that the Mott lobe with $\rho=2$ decreases as $t/W$ grows, a fact that is maintained for any angle; however, it is clear that as the angle grows 
the statistics favor the localization of the particles, which is reflected in the increase of the Mott lobe area and the displacement of the critical 
point towards larger values. The phase diagram shows a much smaller $\rho=2$ Mott lobe than the anyon chain with two-body interactions.
Also, we obtained a re-entrant behavior for the $\rho=2$ Mott lobe regardless of the statistical angle, i. e. for a fixed 
chemical potential value the ground state passes from a Mott insulator to a superfluid phase and then returns to the Mott insulator one. Note that this 
does not happen for the $\rho=1$ Mott lobe (see Fig. \ref{fig9}). The emergence and evolution of the $\rho=1$ Mott lobe as the hopping parameter 
increases is shown Fig. \ref{fig9} for two different angles $\theta=3\pi/4$ and $\theta=\pi$, where a zoom in has been made. From this it is clear 
that if the statistics increase, the localization grows. Note that for negative constant values of the chemical potential, the ground state is a 
superfluid with density lower than one. As the kinetic energy increases, it goes to a Mott insulator state with $\rho=1$ and finally returns to a 
superfluid state, but with a density greater than one. From this figure, it is clear that the Mott lobe area increases with the hopping.\par 
\begin{figure}[t!]
\setlength{\fboxrule}{0.47 pt}
\includegraphics [width=0.47\textwidth]{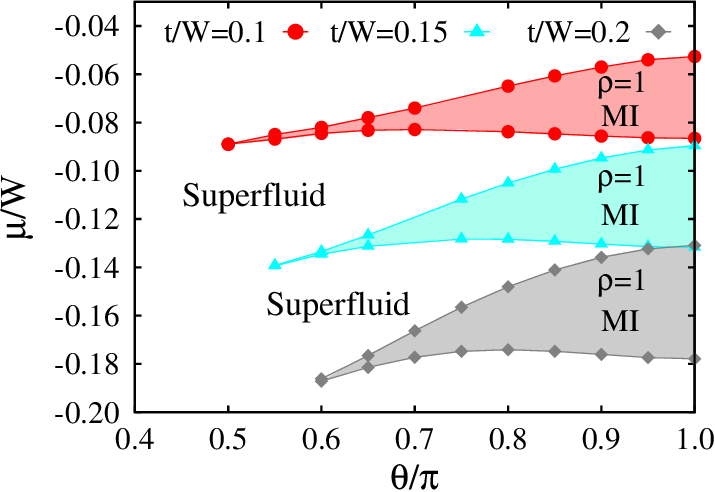} 
\caption{Phase diagram of the anyon-Hubbard model with local three-body interactions in the plane $(\mu/W, \theta/\pi)$. 
We consider three different hopping parameters $t/W=0.1, 0.15$ and $0.2$. The points are extrapolations to the thermodynamic limit from DMRG data 
and the lines are visual guides. MI means Mott insulator regions.}
\label{fig10}
\end{figure}
We showed that the statistical angle drives a quantum phase transition from a superfluid state to a Mott insulator one in the range of 
$\theta=\pi/4$-$3\pi/4$ for a fixed hopping parameter $t/W=0.1$ and global density $\rho=1$ (see Fig. \ref{fig7}). However a pending task is to 
explore the ground state phase diagram as a function of the statistical angle, which is shown in Fig. \ref{fig10} for three different values of the 
hopping parameter, $t/W=0.1$, $0.15$, and $0.20$. We obtain that there is a critical value of the angle at which the Mott insulator phase appears, and 
this critical value moves to greater values as the hopping grows. A reentrance phase transition was observed, since the hole excitation energy curve 
exhibits a maximum. Hence at some suitable constant chemical potential, the model displays a sequence of quantum phase transitions between the Mott 
insulator and the superfluid phases. In the sequence (from left to right), we have a change from a superfluid region with $\rho>1$ to a Mott insulator 
one with $\rho=1$  and then again to an superfluid region ($\rho<1$), and  finally the system remains in an insulator state. Finally, we note that 
for a fixed angle, the Mott lobe area increases with the tunneling. Note that the shape of the Mott lobes around the critical point is not elongated 
like others shown in Fig. \ref{fig2} or others reported  previously, which can indicate that this quantum phase transition is not of the 
Kosterlitz-Thouless type.\par 
It is well known in the literature that using the gap vanishing point to determine the critical point related to a quantum phase transition gives us 
poor results and that some measures of the entanglement can help us in this task. Precisely, it has been shown that for models like the Bose-Hubbard 
one, the estimator based on the von Neumann block entropy $\Delta S_{LK}(L)= S_L(L/2)-S_{L/2}(L/4)$  proposed by L\"auchli and Kollath leads to better 
results~\cite{Lauchli-JSM08}. According to above definition and the expression Eq. (\ref{lauchhh}), we obtain: 
\begin{equation}
   \Delta S_{LK}(L)= \left\{ \begin{array}{lcc}
             \frac{c}{6}ln [2], & \theta \leq \theta_c, \\
             \ 0, & \theta > \theta_c,  
             \end{array}   
\right.
\label{stimator1}
\end{equation}
\noindent $\theta_c$ being the critical angle.\par
\begin{figure}[t!]
  \centering
\setlength{\fboxrule}{0.5 pt}
 \includegraphics [width=0.51\textwidth]{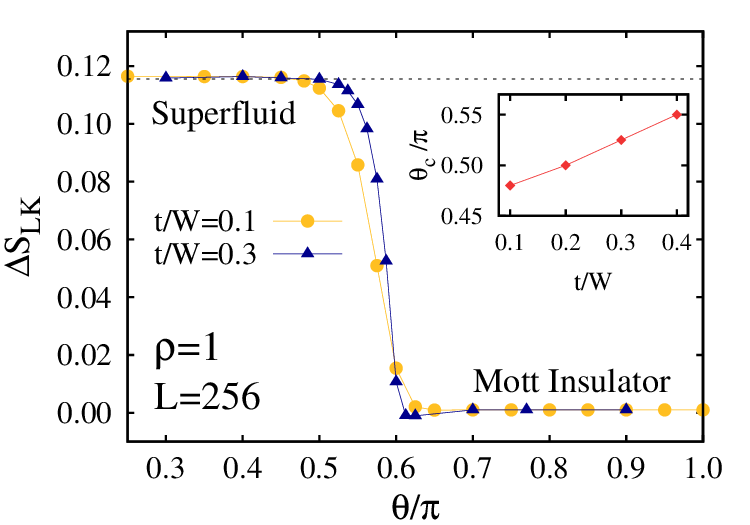} 
 \caption{ The estimator $\Delta S_{LK}$ as a function of angle $\theta$ for $t/W=0.1$ and $t/W=0.3$. Here we fixed L=256 
 and $\rho=1$. The points are DMRG data and the lines are a visual guide. Inset: Evolution of the critical angle $\theta_c$ with the hopping 
 parameter.}
\label{fig11}
\end{figure}
In Fig. \ref{fig11}, the evolution of the estimator $\Delta S_{LK}$ with the statistical angle is shown for an anyon chain with $\rho=1$, and $L=256$.
When $\theta=0$, the ground state is a superfluid, because the $\rho=1$ Mott insulator lobe does not exist for bosons under local three-body 
interactions. Therefore, the estimator will be equal to $ln (2)/6$, and it remains at this point for non-zero values of the statistical angle, 
indicating that there is a range of values of $\theta$ for which  the ground state is a superfluid. However, from a given critical angle the estimator 
collapses to zero within a short range. After that, the estimator remains constant at zero, indicating that the ground state is now a Mott insulator 
one. Regardless of the hopping parameter, we see that the estimator clearly shows us the quantum phase transition from a superfluid to a Mott 
insulator state. Note that the superfluid region becomes greater and the shape of the curve becomes sharper as the hopping increases, although at the 
thermodynamic limit the shape of the curve will be a step function, according to the expression Eq. (\ref{stimator1}). The critical point corresponds 
to the greatest value of $\theta$, for which the estimator is equal to $ln (2)/6$, and we obtain that the position of the critical angle moves to 
greater values as the hopping grows (see inset of Fig. \ref{fig11}). We use the above procedure to determinate the critical points due to we do not 
know \textit{a priori} the order of the transition and that the phase diagram show in Fig. \ref{fig10} suggest that the transition is not the 
Kosterlitz-Thouless type, which prevents us from using the gap-scaling analyses, which led to the most accurate determination of the critical point 
in Bose-Hubbard model~\cite{Carrasquilla-PRA13,Dalmonte-PRB15}. An alternative way to estimate the critical points is calculating the 
Luttinger-liquid parameter, which is a very expensive numerical way to determinate the critical points~\cite{Clay-PRB99}.\par
\begin{figure}[t!]
\centering
\setlength{\fboxrule}{0.47 pt}
\includegraphics [width=0.5\textwidth]{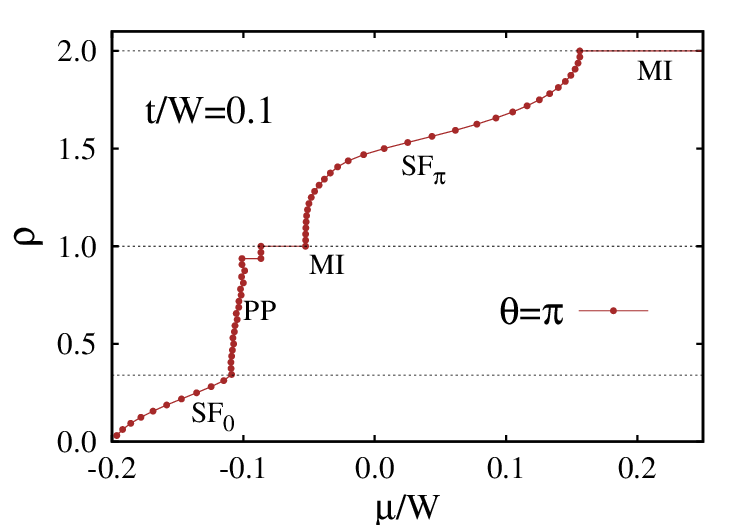} 
\caption{Equation of state $\rho=\rho(\mu)$ for anyons under a finite three-body interaction $t/W=0.1$ in the pseudo-fermion limit ($\theta=\pi$). The 
points are DMRG data and the lines are a visual guide.}
\label{fig12}
\end{figure}
The pseudo-fermion limit ($\theta=\pi$) of the anyon-Hubbard model attracted attention and is full of surprises, for instance Greschner and Santos 
found a Mott insulator phase with one particle per site in absence of interactions and diverse superfluid phases for non-integer 
densities~\cite{Greschner-PRL15}. Recently Zhang and co-workers considered a non-zero finite two-body interaction between the anyons and observed 
that the ground state for non-integer densities is very rich, showing transitions from superfluids to a separation phase or paired phase for densities 
lower or greather than $\rho=1$, respectively~\cite{Zhang-PRA17}. Motivate by these results, we show in Fig. \ref{fig12} the equation of state 
$\rho=\rho(\mu)$ at the pseudo-fermion limit for a finite three-body interaction $t/W=0.1$. For densities lower than $\rho=0.34$, we obtained a 
superfluid phase $SF_0$, after that the scope of the curve changes and a paired phase appears, result that diverge of the result for finite two-body 
interactions where a separation phase was found, this divergence is expected since the effective interaction (statistics plus three-body potential) 
is weaker in our case than in the two-body calculation. Between the Mott insulator plateaus, we obtained a superfluid phase $SF_{\pi}$ in the whole 
region, whereas for a finite two-body interaction a quantum phase transition from a superfluid phase $SF_{\pi}$ to paired phase was reported. We see 
that our results for a finite three-body interaction are more closer to the non-interacting case than the finite two-body interactions, due to the 
smaller effective interactions.\par 

\subsection{\label{iv} Conclusions}

We studied the interplay of multi-body interactions and the statistics in an anyon-Hubbard model, which was mapped to a modified Bose-Hubbard model 
by means of a fractional Jordan-Wigner transformation. This last-named model was explored using the density matrix renormalization group method, and we 
found that the model exhibits gapless and gapped regions in the phase diagram. The expression $\mu/W= \rho -1$ determines the width of the 
Mott lobes at the atomic limit, regardless of the statistics. As in the spinless and spinor boson cases, we obtain that 
for small statistical angles there is no Mott insulator region with density $\rho=1$ when local three-body interactions are considered. However 
the above Mott insulator state emerges as the statistics grow, from a critical value of the local interaction, which is $W/t=1.1$ for $\theta=3\pi/4$,
this fact being one of the main findings of this study.\par
Away from the extreme limits, we found that together the statistics and the local three-body interaction drive a quantum phase transition from a 
gapless phase to a Mott insulator one when the density is $\rho=1$. This quantum phase transition was clearly identified by means of the von Neumann 
block entropy and the estimator proposed by L\"auchli and Kollath~\cite{Lauchli-JSM08}. Using the latter to estimate the critical angles, we see that 
the hopping parameter moves the critical angles to greater values.\par
Although a complete study of the ground state for non-integer densities is out scope of this paper, we found that the equation of state 
$\rho=\rho(\mu)$ at the pseudo-fermion limit for a finite three-body interaction is closer to the non-interacting case, due to the weaker effective 
interactions. Away from the pseudo-fermion limit, we observe diverse gapless phases, such as: superfluid $SF_{0}$, paired phase, superfluid 
$SF_{\pi}$ and a possible separation phase. The quantum phase transitions between the above gapless phases and their evolution with the local 
three-body interactions for different statistical angles will be explore in a future study.

\section*{Acknowledgments}
We thank Sebastian Greschner for useful comments. All authors are thankful for the support of DIEB- Universidad Nacional de Colombia and 
COLCIENCIAS (grant No. FP44842-057-2015). Silva-Valencia and Franco are grateful for the hospitality of the ICTP, where part of this work was done.


\bibliography{Bibliografia}

\end{document}